\documentclass[aps,prb,preprint,superscriptaddress,amsmath,amssymb,floatfix]{revtex4-2}
\usepackage{graphicx}
\usepackage{booktabs}
\usepackage{hyperref}

\date{\today}

\begin{document}
	
	\title[Enantioselective adsorption on magnetic surfaces]{Enantioselective adsorption on magnetic surfaces}
	
	\author{Mohammad Reza Safari}\email{m.safari@fz-juelich.de}
	\affiliation{Peter Grünberg Institute, Electronic Properties (PGI-6), Forschungszentrum Jülich, 52425, Jülich, Germany}
	\affiliation{Jülich Aachen Research Alliance (JARA-FIT), Fundamentals of Future Information Technology, Forschungszentrum Jülich, 52425, Jülich, Germany}
	
	\author{Frank Matthes}
	\affiliation{Peter Grünberg Institute, Electronic Properties (PGI-6), Forschungszentrum Jülich, 52425, Jülich, Germany}
	\affiliation{Jülich Aachen Research Alliance (JARA-FIT), Fundamentals of Future Information Technology, Forschungszentrum Jülich, 52425, Jülich, Germany}
	
	\author{Vasile Caciuc}
	\affiliation{Peter Grünberg Institute and Institute for Advanced Simulation, Quantum Theory of Materials (PGI-1/IAS-1), Forschungszentrum Jülich, 52425, Jülich, Germany}
	
	\author{Nicolae Atodiresei}
	\affiliation{Peter Grünberg Institute and Institute for Advanced Simulation, Quantum Theory of Materials (PGI-1/IAS-1), Forschungszentrum Jülich, 52425, Jülich, Germany}
	
	\author{Claus M. Schneider}
	\affiliation{Peter Grünberg Institute, Electronic Properties (PGI-6), Forschungszentrum Jülich, 52425, Jülich, Germany}
	\affiliation{Jülich Aachen Research Alliance (JARA-FIT), Fundamentals of Future Information Technology, Forschungszentrum Jülich, 52425, Jülich, Germany}
	\affiliation{Fakultät für Physik, Universität Duisburg-Essen, 47057, Duisburg, Germany}
	
	\author{Karl-Heinz Ernst}\email{karl-heinz.ernst@empa.ch}
	\affiliation{Molecular Surface Science Group, Empa, Swiss Federal Laboratories for Materials Science and Technology, 8600, Dübendorf, Switzerland}
	\affiliation{Nanosurf Laboratory, Institute of Physics, The Czech Academy of Sciences, 16200, Prague, Czech Republic}
	\affiliation{Institut für Chemie, Universität Zürich, 8057, Zürich, Switzerland}
	
	\author{Daniel E. Bürgler}\email{d.buergler@fz-juelich.de}
	\affiliation{Peter Grünberg Institute, Electronic Properties (PGI-6), Forschungszentrum Jülich, 52425, Jülich, Germany}
	\affiliation{Jülich Aachen Research Alliance (JARA-FIT), Fundamentals of Future Information Technology, Forschungszentrum Jülich, 52425, Jülich, Germany}
	
	\begin{abstract}
		From the beginning of molecular theory, the interplay of chirality and magnetism has intrigued scientists. There is still the question if enantiospecific adsorption of chiral molecules occurs on magnetic surfaces. Enantiomer discrimination was conjectured to arise from chirality-induced spin separation within the molecules and exchange interaction with the substrate's magnetization. Here we show that single helical aromatic hydrocarbons undergo enantioselective adsorption on ferromagnetic cobalt surfaces. Spin and chirality sensitive scanning tunneling microscopy reveals that molecules of opposite handedness prefer adsorption onto cobalt islands with opposite out-of-plane magnetization. As mobility ceases in the final chemisorbed state, it is concluded that enantioselection must occur in a physisorbed transient precursor state. State-of-the-art spin-resolved ab initio simulations support this scenario by refuting enantio-dependent chemisorption energies. These findings demonstrate that van der Waals interaction should also include spin-fluctuations which are crucial for molecular magnetochiral processes.
	\end{abstract}
	
	
	\maketitle
	\newpage
	
	\section*{Introduction}
	\label{s_main}
	Soon after his seminal discovery of molecular chirality \cite{Pasteur1848} and its common occurrence in organic matter Pasteur conjectured physical fields as its origin \cite{Pasteur1860}. As natural optical activity and Faraday rotation are both manifested by rotating the plane of polarization of light, Pasteur assumed that magnetic fields must be the source of chirality in the universe \cite{Pasteur1884}.  However, Kelvin made clear that magnetic rotation (i.e.\,Faraday rotation) has no chirality \cite{Thomson1894}.  In 1894, Pierre Curie proposed that parallel and antiparallel alignments of electric and magnetic fields will induce chirality \cite{Curie1894}, but as Barron pointed out, such chiral influence will vanish under conditions of thermodynamic equilibrium \cite{Barron1994}.  However, it was shown later that collinear aligned light and magnetic field are a truly chiral influence \cite{Wagniere1982,Rikken2000}. 
	
	Recent experiments have demonstrated an interaction between electron spin and molecular chirality, coined as chirality-induced spin selectivity (CISS) \cite{Ray1999,Rosenberg2011, Naaman2012}. The characteristic feature of CISS is the spin-dependent propagation of electrons in chiral molecules, as evidenced by photoemission \cite{Ray1999, Gohler2011a, Kettner2018a} and electric transport \cite{xie11-00,kira16-00,arag17-00,Al-Bustami2020,Mishra2020} experiments. However, profound understanding of CISS including a suitable theory framework is still lacking \cite{EVER22-00}.
	
	Interestingly, recent reports of enantioselective adsorption on magnetized substrates were also considered a manifestation of CISS \cite{Banerjee-Ghosh2018c,Tassinari2019,Lu2021,Nguyen2022}. In all cases, evidence of enantiospecificity was provided indirectly, either by temporal separation under flow conditions from solution onto ferromagnetic substrates coated with 5 to 10\,nm Au for oxidation protection or integrally over large ensembles. However, no direct evidence of lateral separation depending on absolute handedness and magnetization has been provided so far. In particular, no detailed information on the structure and orientation of the adsorbate were provided, which makes theoretical approaches extremely difficult, if not impossible. In recent studies of chiral molecules on ferromagnetic Ni(100), neither enantiomeric separation nor differences in adsorption kinetics between magnetic domains were observed \cite{Radetic2022,balj21-00}.
	
	Here we report enantiospecific adsorption of heptahelicene, a chiral helical polyaromatic molecule, on ferromagnetic single-crystal cobalt surfaces in ultrahigh vacuum by means of spin-polarized scanning tunneling microscopy (SP-STM). Lateral enantioselection onto domains of opposite out-of-plane magnetization occurs in a transient physisorbed precursor state before final chemisorption. Density functional theory (DFT) calculations show degenerated ground states in the chemisorbed state of the enantiomers on the ferromagnetic surface, suggesting that enantiospecificity arises from spin-dependent van der Waals (vdW) interactions.
	
	\section*{Determination of molecular handedness and substrate magnetization}
\begin{figure*}[t!]%
		\centering
		\includegraphics[width=\textwidth]{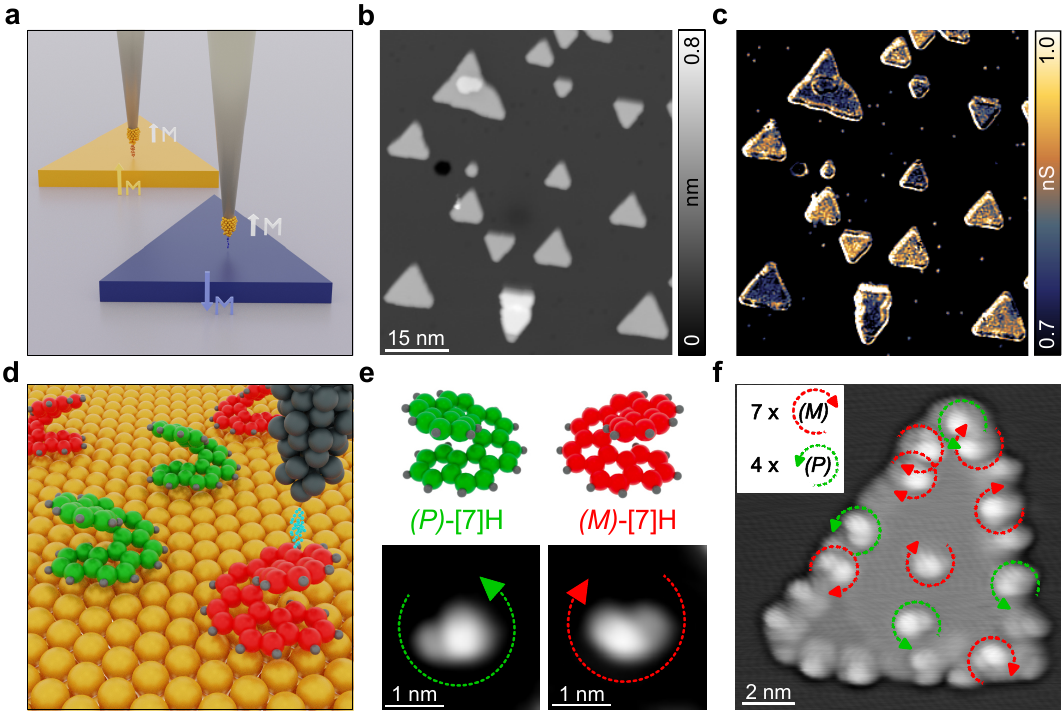}
		\caption{\textbf{$\mid$ Principles of spin-polarized and enantio-resolved STM.} 
			\textbf{a}, Sketch of two oppositely out-of-plane magnetized Co nanoislands probed with a magnetic Co-functionalized STM tip. 
			\textbf{b}, Constant-current topographic STM image of triangular Co nanoislands on Cu(111). The two triangle orientations are due to different stacking sequences of the Co layers on the Cu surface.
			\textbf{c}, $dI/dV$ map at $V_{\rm bias}=-600$\,mV measured with a magnetic Co-functionalized STM tip simultaneously with the topographic image in \textbf{b}. 
			\textbf{d}, Sketch of STM imaging of single molecules. 
			\textbf{e}, Ball-and-stick model of [7]H enantiomers and assignment of their absolute handedness from topographic STM contrast. Counterclockwise increase of brightness denotes a {\it (P)}-enantiomer (left), while clockwise increase of brightness denotes an {\it (M)}-enantiomer (right).
			\textbf{f}, Example for 'chirality counting' of [7]H molecules on a single Co nanoisland. Indicated by circular arrows, 4 {\it (P)}- and 7 {\it (M)}-enantiomers are identified.
			STM parameters: \textbf{b,c} $V_{\rm bias}=-600$\,mV, $I_{\rm t}=1$\,nA, $V_{\rm mod}=10$\,mV, $f_{\rm mod}=875$\,Hz, 5\,K, Co-functionalized W tip, \textbf{e} $V_{\rm bias}=1$\,V, $I_{\rm t}=50$\,pA, 5\,K, W tip, \textbf{f} $V_{\rm bias}=1$\,V, $I_{\rm t}=50$\,pA, 5\,K,  Gaussian high-pass filtered.}
		\label{principles}
	\end{figure*}

	Bilayer Co nanoislands are created on a Cu(111) single-crystal surface by evaporation of metallic cobalt \cite{DeLaFiguera1993,Safari2022}. Such islands are known to exhibit out-of-plane magnetization \cite{piet04-00,piet06-00,oka10-00,park17-00,Esat2017,metz21-00}. The magnetization directions of the islands are probed by spin-polarized scanning tunneling microscopy and spectroscopy (SP-STM/STS). That is, the tungsten tip of the STM is modified with Co prior to imaging (see Methods for procedure). A contrast in differential conductance ($dI/dV$) arises from different spin-polarized tunneling probabilities between the oppositely magnetized Co nanoislands and the magnetic STM tip \cite{piet04-00,piet06-00,oka10-00,park17-00,Esat2017,metz21-00} (Figs.\,\ref{principles}a-c). With the absolute magnetization direction of the STM tip unknown, there is no unique assignment of higher or lower $dI/dV$ signal to the absolute direction of the island magnetization $\boldsymbol{M}$. For the sake of analysis, a higher $dI/dV$ signal (yellow contrast) is arbitrarily assigned as up magnetization ($\boldsymbol{M}$ pointing to the vacuum; $\boldsymbol{M}=\,\uparrow$) and a lower $dI/dV$ signal (blue contrast) as down magnetization ($\boldsymbol{M}$ pointing to the copper substrate; $\boldsymbol{M}=\,\downarrow$).
	
	Sub-monolayer coverages of chiral heptahelicene C$_{30}$H$_{18}$ molecules ([7]H) are deposited in ultrahigh vacuum by sublimation of racemic powder onto these ferromagnetic islands. The handedness of the molecules is then determined for each adsorbate one-by-one using STM (Fig.\,\ref{principles}d). As the [7]H molecules are adsorbed with their proximal phenanthrene group parallel to the surface (see DFT results in Supplementary Section\,S7), their helical axis is oriented perpendicular to the surface. The absolute handedness is therefore clearly distinguishable from a clockwise or counterclockwise increase in apparent height in constant-current STM images (Fig.\,\ref{principles}e and Supplementary Fig.\,S9c-f). For each island, the ratio of left- to right-handed molecules adsorbed on it is then determined in order to analyze any enantiospecific interaction (Fig.\,\ref{principles}f).
	
	\section*{Magneto-enantiospecific [7]H adsorption on Co}
	\label{sec_results}
	
\begin{figure*}[b]%
		\centering
		\includegraphics[width=0.85\textwidth]{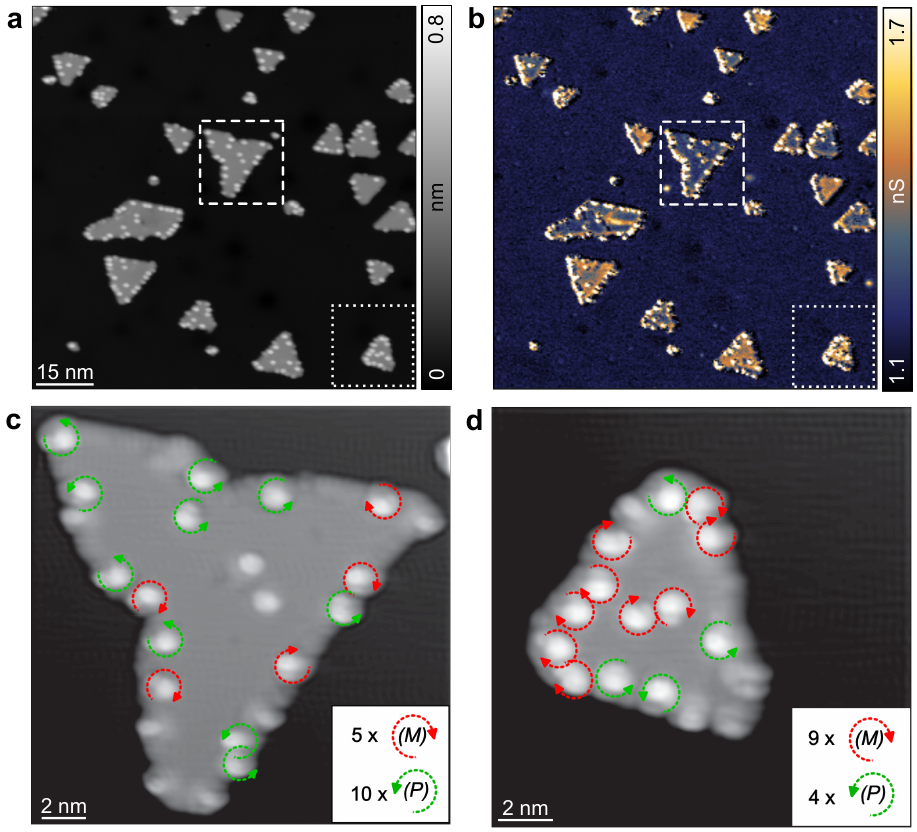}
		\caption{\textbf{$\mid$ Enantiospecific adsorption of [7]H on ferromagnetic Co nanoislands.} 
			Topographic images reveal the handedness of each molecule and the spectroscopic maps allow to distinguish between the opposite magnetization directions of the Co islands. Both quantities are determined individually for each molecule. \textbf{a}, Constant-current topographic image showing [7]H molecules attached only to Co nanoislands. 
			\textbf{b}, Simultaneously with \textbf{a} measured differential conductance ($dI/dV$) map scaled in nS revealing magnetic contrast (blue versus yellowish) between Co nanoislands with opposite out-of-plane magnetization ($V_{\rm bias}=-600$\,mV, $I_{\rm t}=950$\,pA, $V_{\rm mod}=20$\,mV, $f_{\rm mod}=752$\,Hz, 5\,K, Co-functionalized W tip).
			\textbf{c} and \textbf{d}, High-resolution topographic STM images of the areas framed in \textbf{a} and \textbf{b} by dashed and dotted lines, respectively ($V_{\rm bias}=1000$\,mV, $I_{\rm t}=50$\,pA, 5\,K,  Gaussian high-pass filtered). Green and red circular arrows indicate the handedness of the respective [7]H molecules. The insets represent the number of right-handed {\it (P)} and left-handed {\it (M)} molecules adsorbed on the corresponding island.}
		\label{heli_mag_Co}
	\end{figure*}

	In principle, the adsorption of [7]H occurs in four possible combinations of substrate magnetization direction ($\boldsymbol{M}=\,\uparrow$ or $\boldsymbol{M}=\,\downarrow$) and enantiomer handedness [{\it (P)} or {\it (M)}]. Enantiospecific adsorption of [7]H molecules on Co islands is evidenced by directly counting the occurrence of the four combinations in topographic STM images measured on Co nanoisland with magnetization directions determined from spin-polarized differential conductance maps. Figure\,\ref{heli_mag_Co} presents results of such a chirality counting procedure in one STM frame. A constant-current topographic STM image shows [7]H molecules on Co/Cu(111) at 5\,K after deposition at room temperature (RT), see Fig.\,\ref{heli_mag_Co}a. At the low coverages investigated here, the triangular Co nanoislands are decorated with molecules, while the Cu substrate remains bare. Hence, the surface mobility of the [7]H molecules during deposition allows them to travel to the Co nanoislands, where they accumulate as a result of stronger binding \cite{Jia-2018}. The preferential adsorption of the molecules at the rims of the islands is analyzed and discussed in Section\,S8 of the Supporting Information. The $dI/dV$ map reveals magnetic contrast of the Co nanoislands (Fig.\,\ref{heli_mag_Co}b), demonstrating Co island ferromagnetism also in the presence of chemisorbed [7]H molecules \cite{Safari2022}. Figures\,\ref{heli_mag_Co}c and d present high-resolution topographic images of the dashed and dotted framed areas in Figs.\,\ref{heli_mag_Co}a and b. The two islands chosen have opposite magnetization, i.e.,\,$\boldsymbol{M}=\,\downarrow$ and $\boldsymbol{M}=\,\uparrow$. The absolute handedness of the helicene molecules is indicated by red and green circular arrows for {\it (M)}-[7]H and {\it (P)}-[7]H molecules, respectively. The arrows spiral upward from the proximal to the distal end of the molecular helix. The enantiomer count on both islands yields opposite enantiomeric imbalances, i.e., {\it (M)}/{\it (P)} ratios of 5/10 (Fig.\,\ref{heli_mag_Co}c) and 9/4 (Fig.\,\ref{heli_mag_Co}d), which suggests indeed magneto-enantiospecific adsorption of [7]H on ferromagnetic Co islands (for more examples of STM images resolving the molecular handedness, see Supplementary Fig.\,S4).
	
	\begin{figure*}[t]%
		\centering
		\includegraphics[width=1.0\textwidth]{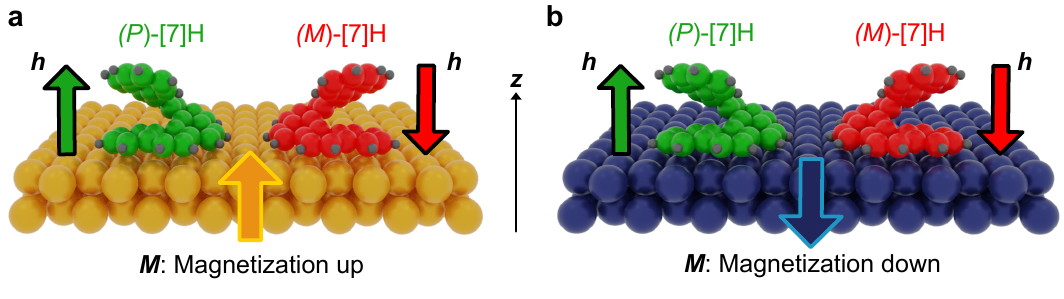}
		\caption{\textbf{$\mid$ Adsorption configurations of [7]H on Co nanoislands} Both enantiomers, {\it (P)}-[7]H and {\it (M)}-[7]H, can adsorb on Co nanoislands with both (\textbf{a}) magnetization up ($\boldsymbol{M}=\uparrow$) or (\textbf{b}) magnetization down ($\boldsymbol{M}=\downarrow$). The pitch vectors of the molecular helices $\boldsymbol{h}=\uparrow$ (green) and $\boldsymbol{h}=\downarrow$ (red) represent {\it (P)}-[7]H and {\it (M)}-[7]H enantiomers, respectively (see Supplementary Section\,S1).}
		\label{scheme}
	\end{figure*}
	
	To support this observation with better statistics, two independent data sets were measured using separately prepared Co/Cu(111) substrate systems, different runs of [7]H deposition, and newly Co-functionalized STM tips (see Supplementary Section\,S2). 
	For the sake of further statistical analyses, the enantiomer handedness is expressed here as 'pitch vector' $\boldsymbol{h}$.
	$\boldsymbol{h}=\uparrow$ represents the {\it (P)}-[7]H enantiomer and $\boldsymbol{h}=\downarrow$ the {\it (M)}-[7]H enantiomer (Fig.\,\ref{scheme}). 
	A more detailed definition of $\boldsymbol{h}$ is given in Supplementary Section\,S1. 
	The number of occupation for each of the four possible adsorption configurations are denoted by $N_{\boldsymbol{M}\boldsymbol{h}}$, with $\boldsymbol{M}=\,\uparrow$ or $\boldsymbol{M}=\downarrow$ and $\boldsymbol{h}=\,\uparrow$ or $\boldsymbol{h}=\downarrow$. Hence,
	$\boldsymbol{M}$ and $\boldsymbol{h}$ can be parallel ($N_{\uparrow\uparrow}$ and $N_{\downarrow\downarrow}$) or antiparallel ($N_{\uparrow\downarrow}$ and $N_{\downarrow\uparrow}$), see Fig.\,\ref{scheme}.
	
	\begin{table*}[b!]
		\begin{center}
			\begin{minipage}{\textwidth}
				\caption{$\mid$ \textbf{Enantioselective counting of [7]H on Co nanoislands.} Occupation numbers $N_{\boldsymbol{M}\boldsymbol{h}}$ of the four combinations of $\boldsymbol{M}$ and $\boldsymbol{h}$ in the data presented in Fig.\,\ref{heli_mag_Co} and Supplementary Sec.\,S2. The number of molecules for which the handedness could not be determined and the total number of analyzed molecules are also listed.}\label{table_exp}
				\begin{tabular*}{\textwidth}{@{\extracolsep{\fill}}lcccccc@{\extracolsep{\fill}}}
					\toprule
					Data sets& $N_{\uparrow\uparrow}$ & $N_{\uparrow\downarrow}$ & $N_{\downarrow\uparrow}$ & $N_{\downarrow\downarrow}$ & Indeterminable  &  Analyzed  \\
					&  &  &  &  &  handedness &   molecules \\
					\midrule
					Set 1  & 39 & 61 & 98 & 78 & 15 & 291 \\
					Set 2  & 63 & 94 & 172 & 114  & 14 & 457 \\
					Total  & 102 & 155 & 270 & 192 & 29  & 748 \\
					\botrule
				\end{tabular*}
			\end{minipage}
		\end{center}
	\end{table*}
	
	Statistical analysis of more than 740 molecules on 107 islands revealed a pronounced imbalance of the handedness of adsorbed [7]H molecules with respect to the direction of the out-of-plane substrate magnetization. Table\,\ref{table_exp} summarizes the statistical counts of two different data sets.
	Dividing the adsorbed molecules into these two classes ($\boldsymbol{M}$ and $\boldsymbol{h}$ parallel or antiparallel, respectively) plus taking the different island areas by a weighting factor ($\alpha={A_{M_\uparrow}}/{A_{M_\downarrow}}$) into account gives: $N_1=N_{\uparrow\uparrow}+\alpha N_{\downarrow\downarrow}$ and $N_2=N_{\uparrow\downarrow}+\alpha N_{\downarrow\uparrow}$. 
	For both data sets, the antiparallel $\boldsymbol{M}$-$\boldsymbol{h}$ alignment outnumbers the parallel alignment ($N_1<N_2$).
	Magneto-enantiospecific adsorption is identified if the ratio of $r={N_1}/{N_2}$ deviates from unity, i.e. $r\neq 1$. 
	The statistical analysis of our data in the framework of a trinomial distribution (see Supplementary Section\,S3) yields for the combined data sets an magneto-enantiospecific ratio $r=0.68\pm 0.06$, i.e.\,$r$ is about five standard deviations lower than one (Table\,\ref{table_ana}).
	Hence, the adsorption probability of [7]H enantiomers on a ferromagnetic Co(111) islands clearly depends on the  molecular handedness and the substrate magnetization, meaning that 
	the {\it (M)}-[7]H enantiomer is preferably found on Co islands with one magnetization direction (e.g. $\boldsymbol{M}=\uparrow$), whereas the {\it (P)}-[7]H enantiomer preferably adsorbs on nanoislands with the opposite magnetization direction (e.g. $\boldsymbol{M}=\downarrow$).

	\begin{table*}[htb]
		\begin{center}
			\begin{minipage}{\textwidth}
				\caption{\textbf{$\mid$ Statistics of [7]H adsorption on Co nanoislands.} 
					$\alpha={A_{M_\uparrow}}/{A_{M_\downarrow}}$ is the ratio of the total areas of islands with $\boldsymbol{M}=\uparrow$ and $\boldsymbol{M}=\downarrow$, respectively.
					$N_1=N_{\uparrow\uparrow}+\alpha N_{\downarrow\downarrow}$ and $N_2=N_{\uparrow\downarrow}+\alpha N_{\downarrow\uparrow}$ are the number of molecules in classes 1 and 2. 
					$N_3$ is the number of molecules with indeterminable handedness.  
					The error bar of $r$ is calculated according to Equation\,(S1) in Supplementary Section\,S3. 
					$\Delta E$ is the energy difference of the two molecule classes, which is derived from $r$ using Equation\,(\ref{equation_Boltzmann}).}
				\label{table_ana}
				\begin{tabular*}{\textwidth}{@{\extracolsep{\fill}}lcccccc@{\extracolsep{\fill}}}
					\toprule%
					Data & $\alpha=$ & 
					$N_1=$ & $N_2=$ & 
					$N_3$ & $r=$ & $\Delta E$\\
					sets & ${A_{M_\uparrow}}/{A_{M_\downarrow}}$ & 
					$N_{\uparrow\uparrow}+\alpha N_{\downarrow\downarrow}$ & $N_{\uparrow\downarrow}+\alpha N_{\downarrow\uparrow}$ & & $N_1/N_2$ & (meV) \\
					\midrule
					Set 1  & 0.62 &  88 & 122 & 15 & $0.72\pm 0.10 $ & $ 8 \pm 4 $ \\
					Set 2  & 0.49 & 119 & 178 & 14 & $0.67\pm 0.08 $ & $10 \pm 3 $ \\
					Total  & 0.55 & 207 & 303 & 29 & $0.68\pm 0.06 $ & $10 \pm 2 $ \\
					\botrule
				\end{tabular*}
			\end{minipage}
		\end{center}
	\end{table*}

	\section*{Precursor-mediated chemisorption at RT}
	\begin{figure*}[t!]%
		\centering
		\includegraphics[width=\textwidth]{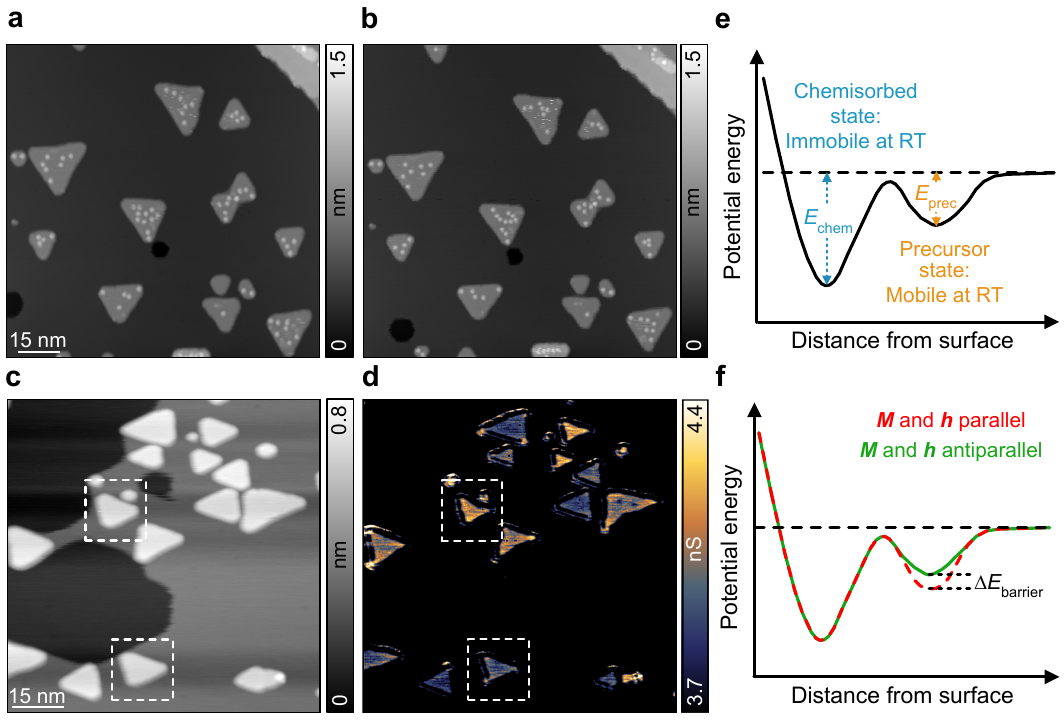}
		\caption{\textbf{$\mid$ Magneto-enantiospecific [7]H adsorption at RT}. 
			\textbf{a} Constant-current STM images of [7]H molecules deposited on Co nanoislands on Cu(111) at RT and imaged at 296\,K after right after deposition. 
			\textbf{b}, STM image of same area under same conditions as in \textbf{a} imaged after 329\,min. Only small lateral movement of molecules is observed, and all molecules remained on their island ($V_{\rm bias}=500$\,mV, $I_{\rm t}=20$\,pA, 296\,K, W tip). 
			\textbf{c}, Constant-current topographic image of Co nanoislands on Cu(111) deposited and measured at RT. Dashed squares indicate two islands of approximately equal size and identical stacking.
			\textbf{d}, Simultaneously measured $dI/dV$ map of the same area as in \textbf{c}. The different contrast of quasi-identical islands (dashed squares) is assigned to ferromagnetism ($V_{\rm bias}=-600$\,mV, $I_{\rm t}=250$\,pA, $V_{\rm mod}=20$\,mV, $f_{\rm mod}=752$\,Hz, 296\,K, Co-functionalized W tip).
			\textbf{e}, Sketch of the potential energy curve of adsorption. Prior to reaching a strongly bound chemisorbed state, the molecule can travel laterally long distances in a physisorbed precursor state. 
			\textbf{f}, Sketch of magnetochiral binding differences in the physisorbed precursor state. 
		}
		\label{RT_data}
	\end{figure*}
	[7]H molecules sublimed from a racemate arrive on the sample surface at random positions and with uniformly distributed handedness. Hence, the observed adsorption preference based on island magnetization and molecular handedness reflects a process in which the molecules can travel across the surface in order to choose islands with preferred magnetization. However, STM imaging at, RT at which deposition is performed, shows clearly that molecules have very limited mobility once chemisorbed on an island, thus leaving the Cu regions between the Co islands free of molecules and suppressing exchange of molecules between islands (Figs.\,\ref{RT_data}a and b and Supplementary Section\,S4).
	
	The finding that the magnetization-dependent enantiomeric excess on the Co nanoislands develops during deposition at RT requires that the superparamagnetic blocking temperature of the Co bilayer nanoislands is above RT on the time scale of the precursor-mediated adsorption process. Otherwise, the island magnetizations $\boldsymbol{M}$ would fluctuate thermally and could not cause the enantiomeric excess on the islands. Such a conclusion is experimentally verified by performing spin-polarized measurements at RT (for details see Supplementary Section\,S5), which revealed clear and stable magnetic contrast of the Co nanoislands (Fig.\,\ref{RT_data}c and d). 
	
	We also examined our data to determine whether the observed enantiospecificity could be due to other properties of the Co nanoislands. Separate evaluation of the enantiomeric ratio for [7]H molecules at the rim and center of the islands revealed stronger enantioselectivity at the center than at the rim. This allows us to rule out scenarios in which specific electronic properties of the rim area \cite{piet06-00} or the diffusion barrier of the molecules to overcome the step at the island edge, e.g.\,at chiral kink sites, play a role in enantiospecific adsorption (for details see Section\,S8, Supporting Information). Similarly, testing the relationship between the enantiomeric excess of the islands and their stacking type (unfaulted or faulted) \cite{negu08-00} rather than magnetization direction revealed no significant effect (for details see Section\,S9, Supporting Information). This finding is corroborated by the experimental observations that (i) magneto-enantiospecific adsorption due to the magnetization direction occurs for both stacking types with equal selectivity and (ii) there is no statistically significant correlation between island stacking and islands magnetization direction, see Sections\,S9 and S10 (Supporting Information).
	
	\section*{DFT results}
	\label{sec_dft}
	
	Important additional insight is provided by spin-polarized first-principles calculations of the [7]H -- 2\,ML Co/Cu(111) system in the framework of dispersion-corrected DFT (see Methods). As already mentioned, the [7]H molecule adsorbs with the proximal phenanthrene group aligned basically parallel to the Co surface (see Supplementary Fig.\,S9a). Twelve carbon atoms of that group are at distances between 210 and 240\,pm to Co atoms. The adsorption energy evaluated from these calculations amounts to 3.51\,eV per molecule, which, together with the short C-Co distances, indicates that the molecule strongly chemisorbs to the ferromagnetic Co substrate. Such strong chemisorption explains in part the low mobility of [7]H at RT. Moreover, no differences in binding energy (within the error of these ab initio calculations) are obtained in non-collinear spin-polarized calculations when the Co magnetization points up or down (see Supplementary Section\,S7 for details). Molecular handedness and magnetically induced asymmetric spin-orbit interactions do not play a role once the molecule reaches the final chemisorption state.
	
	\section*{Discussion}
	\label{sec_discussion}
	
	As no interisland exchange of molecules is possible in the chemisorbed state, magnetochiral selection must occur in a transient precursor state. Such an adsorption precursor state is a physisorbed state that is transiently occupied when molecules approach the surface before reaching the ground state of chemisorption (Fig.\,\ref{RT_data}e). Precursor states were originally introduced in order to explain deviations from Langmuirian adsorption and desorption kinetics \cite{wein87-00,bowk16-00}, and have been directly observed via STM for aromatic compounds \cite{brow98-00}. The larger distance from the surface and a much weaker adsorption energy of molecules in a precursor state result in enhanced lateral mobility and allows molecules to seek out preferred adsorption sites over a large distance \cite{bowk16-00}. 
	
	In general, adsorption may occur under thermodynamic or kinetic control. The ratio of the magnetochiral occupation numbers $N_1$ and $N_2$ can be related via a Boltzmann factors to an energy difference $\Delta E$ by
	\begin{equation}
		r=\frac{N_1}{N_2}=\exp\left(-\frac{\Delta E}{k_{\rm B}T}\right),
		\label{equation_Boltzmann}
	\end{equation}
	see Supplementary Section\,S6. The statistics in Table\,\ref{table_ana} yields $\Delta E = (10\pm2)$\,meV for $T=296$\,K.
	However, as diffusion or desorption from the strongly bound chemisorbed state can be safely excluded, thermodynamic equilibrium is not achieved. Furthermore, our DFT calculations did not yield different chemisorption energies for the enantiomers that could lead to enantiomeric excess in the ground state. The energy difference is therefore based on different transition rates between precursor and chemisorbed state, due to magneto-enantiospecific binding in the precursor state (Fig.\,\ref{RT_data}f and Supplementary Fig.\,S8).
	
	Physisorption in the precursor state is based on dispersive forces and enantioselection in that state points to an electron spin-selective process at the vdW level. 
	Our results seem therefore to support the controversial picture of vdW interactions that go beyond spin averaging but include a spin-dependence in dispersive forces \cite{vydr09-00,lund10-00}. The well-defined [7]H -- 2\,ML Co/Cu(111) system holds the potential to become the drosophila for the development and improvement of novel spin-dependent vdW functionals.
	
	In summary, a pronounced magnetochiral selectivity during adsorption of a helical hydrocarbon on ferromagnetic Co nanoislands has been observed under well-defined conditions in ultrahigh vacuum at the single-molecule level by spin-polarized STM. An influence of collective or ensemble effects on the adsorption behavior can therefore be clearly excluded. The selection occurs in a van der Waals-bound precursor state, allowing the molecules to sample several Co islands prior to final chemisorption. Our report shines new light onto magnetochiral phenomena in molecule-surface interactions and provides a well-defined adsorbate system for theoretical studies.
	
	\section*{Methods}\label{sec_Methods}
	Measurements were performed in an ultrahigh vacuum (UHV, $<10^{-8}$\,Pa) cluster tool equipped with a preparation chamber for substrate preparation and analysis, a separate molecule deposition chamber, and a LT-STM (Omicron Scienta) operating down to 5\,K.
	
	\subsection{Molecule synthesis}
	Racemic mixtures of heptahelicene were synthesized from a stilbene precursor by photocyclyzation as described previously \cite{SUDH86-00}.
	
	\subsection{Sample preparation}
	Co bilayer nanoislands were grown on an atomically clean Cu(111) surface by in-situ deposition at RT from a $99.99 \%$ pure Co rod using e-beam evaporation. The deposition rate was 0.2\,ML/min and pressure during deposition was $<2\times 10^{-8}$\,Pa. The cleaning procedure of the Cu(111) single crystal by repeated cycles of Ar$^+$ sputtering followed by annealing. Further details about the Co deposition can be found in Ref.\,\cite{Safari2022}.
	After Co deposition, the shape and density of the Co bilayer islands as well as the cleanness of the deposition were checked by means of topographic STM images. Simultaneously, spin-polarized STM measurements were performed to confirm the presence of a well-defined magnetization of the islands perpendicular to the surface. Then, the Co/Cu(111) substrate was transferred at RT to the molecule chamber and exposed to the molecular vapour sublimed from a carefully heated glass crucible. The crucible temperature was about 400\,K, the exposure time typically less than one minute, and the pressure during sublimation less than $10^{-7}$\,Pa \cite{Safari2022}.   
	Finally, the sample was transferred to the STM chamber within a few minutes and immediately cooled for the STM experiments, which were performed at a pressure in the mid $10^{-9}$\,Pa range and at 5\,K unless otherwise specified. Higher measurement temperatures $T_{\rm meas}$ were achieved by cooling with liquid N$_2$ instead of liquid He ($T_{\rm meas}=78$\,K), during the slow warming of the system to RT ($T_{\rm meas}=285$\,K), or by measuring immediately after [7]H deposition at RT without any cooling ($T_{\rm meas}=296$\,K).
	
	\subsection{STM measurements}
	Electrochemically etched (polycrystalline W wire in 5\,M NaOH solution) were used for all STM measurements. Co-functionalized W tips providing magnetic contrast were obtained by intentionally bringing the tip in near mechanical contact with a remote ferromagnetic Co nanoisland and by applying short bias voltage pulses to transfer Co atoms from the island to the tip. The functionalization procedure was repeated until successful acquisition of a spin-polarized $dI/dV$ map from Co nanoislands confirmed the sensitivity of the tip to the out-of-plane magnetization component.
	
	We present two types of STM data: (i) Topographic STM images were taken in constant-current mode. The bias voltage $V_{\rm bias}$ was applied to the sample, and $I_{\rm t}$ denotes the setpoint current of the STM feedback loop. (ii) Differential conductance ($dI/dV$) maps that were taken simultaneously with constant-current topographic images, i.e.\,with the STM feedback loop closed. $dI/dV$ maps represent, to a good approximation, the local density of states as a function of the lateral position and at the energy $eV_{\rm bias}$, where $V_{\rm bias}=0$ corresponds to the Fermi energy. The $dI/dV$ signal for the conductance maps was obtained by modulating $V_{\rm bias}$ with a small sinusoidal signal (rms amplitude $V_{\rm mod}=20$\,mV; frequency $f_{\rm mod}=752$\,Hz) and detecting the resulting modulation of the tunneling current at $f_{\rm mod}$ using a lock-in amplifier. 
	
	The high-resolution topographic STM images in Figs.\,\ref{principles}e,f and  \ref{heli_mag_Co}c,d as well as Supplementary Figs.\,S4 and S9c,e were processed with a Gaussian high-pass filter to enhance the contrast of intramolecular fine structure as described in Ref.\,\cite{Safari2022}.
	
	\subsection{DFT calculations}
	We performed spin-polarized first-principles calculations in the framework of density functional theory (DFT) by employing the generalized gradient approximation (PBE) \cite{Perdew1996} and the optB86b non-local van der Waals exchange-correlation functional \cite{Hamada2014} in a projector augmented plane-wave formulation \cite{Bloechl1994} as implemented in the VASP code \cite{Kresse1994,Kresse1996}. 
	
	The [7]H -- 2\,ML Co/Cu(111) system was modeled within the supercell approach [($8\times8$) in-plane surface unit cells and 1.45\,nm vacuum above the molecule, i.e.\,$2.056 \times 2.056 \times 3.100$\,nm$^3$] containing five atomic layers (2\,Co and 3\,Cu) with the adsorbed molecule on one side of the slab \cite{Makov1995}. Using a plane-wave energy cutoff of 500\,eV in our ab-initio calculations, the two Co layers and the molecule atoms are allowed to relax until the atomic forces are lower than 0.05\,eV/nm.
	
	\section*{Acknowledgments}
	We are grateful for stimulating discussions with G.\,Bihlmayer. N.A.\,acknowledges Deutsche Forschungsgemeinschaft (DFG) support through the Collaborative Research Center SFB 1238 Project No.\,277146847 (subproject C01). The authors gratefully acknowledge the computing time granted by the JARA Vergabegremium and provided on the JARA Partition part of the supercomputer JURECA at Forschungszentrum Jülich.
	
	%

\end{document}